\documentclass[iop,showkeys]{emulateapj}

\shorttitle{Testing the WEP with optical Crab pulses}
\shortauthors{Leung et al.}

\usepackage{amsmath,amssymb}

\usepackage{enumitem}
\newlist{indenteddesc}{description}{1}
\setlist[indenteddesc]{
  leftmargin=0.8in,  
  rightmargin=0in,
  labelindent=0in, 
  labelwidth=0.5in,
  labelsep=0.1in}

\usepackage{natbib}

\usepackage{hyperref}
\hypersetup{colorlinks=true,
    linkcolor=blue,
    filecolor=magenta,      
    urlcolor=cyan}
\usepackage{graphicx}
\usepackage{dcolumn}

\usepackage{bm}
\usepackage{hyperref}

\usepackage{savesym}
\savesymbol{tablenum}
\usepackage{siunitx}
\restoresymbol{SIX}{tablenum} 

\usepackage{soul}

\usepackage{xcolor}

\begin{document}

\title{Testing the Weak Equivalence Principle using Optical and Near-Infrared Crab Pulses} 


\author{Calvin Leung, Beili Hu, Sophia Harris, Amy Brown, and Jason Gallicchio}
    \email{cleung@hmc.edu}
    \email{jason@hmc.edu}
\affil{Physics Department, Harvey Mudd College,
    Claremont, CA 91711}
\author{Hien Nguyen}
\affil{Jet Propulsion Laboratory, Pasadena, CA 91109}




\begin{abstract}
The Weak Equivalence Principle states that the geodesics of a test particle in a gravitational field are independent of the particle's constitution. To constrain violations of the Weak Equivalence Principle, we use the one-meter telescope at Table Mountain Observatory near Los Angeles to monitor the relative arrival times of pulses from the Crab Pulsar in the optical ($\lambda \approx \SI{585}{\nm}$) and near-infrared ($\lambda \approx \SI{814}{\nm}$) using an instrument which detects single photons with nanosecond-timing resolution in those two bands. The infrared pulse arrives slightly before the visible pulse. Our three analysis methods give delays with statistical errors of $\Delta t_{obs} = 7.41 \pm 0.58$, $0.4 \pm 3.6$, and $7.35 \pm 4.48$ microseconds (at most 1/4000 of the pulsar period). We attribute this discrepancy to systematic error from the fact that the visible and infrared pulses have slightly different shapes. Whether this delay emerges from the pulsar, is caused by passing through wavelength-dependent media, or is caused by a violation of the equivalence principle, unless there is a fine-tuned cancellation among these, we set the first upper limit on the differential post-Newtonian parameter at these wavelengths of $\Delta \gamma < 1.07 \times 10^{-10}~(3\sigma)$. This result falls in an unexplored region of parameter space and complements existing limits on equivalence-principle violation from fast radio bursts, gamma ray bursts, as well as previous limits from the Crab.

\end{abstract}



\section{Introduction}
%

For decades, precise experimental tests of general relativity have been carried out using a variety of observational and laboratory techniques, ranging from the Weber bars of the 1960s to astrophysical timing measurements, atom interferometric techniques, and Advanced LIGO ~\citep{weber1960detection,shapiro1968fourth,shapiro1976verification,taylor1989further,williams2004progress,dimopoulos2007testing,zhou2015test,abbott2016observation}.
The parametric post-Newtonian (PPN) formalism translates the results of these diverse tests into a common language . A set of dimensionless parameters summarize the observational consequences of various alternative gravitational theories and enable different experiments to constrain new physics in a unified framework. See \cite{will2014confrontation} for a review. For example, a well-known general-relativistic effect occurs when a point particle experiences a time delay falling through a gravitational potential $U(\mathbf{r})$ from point A to B:
$$\delta t_{GR} = \dfrac{2}{c^3}\left|\int_{A}^{B} U(\mathbf{r})~d\mathbf{r}\right|.$$
This effect, called the Shapiro delay~\citep{shapiro1968fourth}, is generalized by the PPN framework by the introduction of the parameter $\gamma$,
$$\delta t = \dfrac{1 + \gamma}{c^3}\left| \int_A^B U(\mathbf{r})~d\mathbf{r}\right|,$$
with general relativity being the special case where $\gamma = 1$. In the PPN framework, the Weak Equivalence Principle (WEP) is equivalent to the statement that $\gamma$ is constant, independent of the constituents of whatever is going from $A$ to $B$ through $U(\mathbf{r})$. In order to constrain WEP-violating theories of gravity, e.g. scalar-tensor theories, gravitational theories with non-symmetric metrics, and non-metric theories, tests of the WEP aim to measure the absolute value of $\gamma$. See~\cite{shapiro1976verification}. 

A way to test the WEP is to measure whether $\gamma$ varies between different choices of test particles. For example, $\gamma$ may potentially take on different values for photons of two different wavelengths $\lambda_1$ and $\lambda_2$. This manifests itself in a differing time-of-flight $\Delta t$ for the two photons along a common trajectory between arbitrary points A and B:
$$\Delta t = (\delta t_1 - \delta t_2) = (\gamma_1 - \gamma_2)\left| \dfrac{\int_{A}^{B} U(\mathbf{r})~d\mathbf{r}}{c^3} \right|$$
A powerful way to set upper limits on this possibility is to make a differential measurement $\gamma_1 - \gamma_2 \equiv \Delta \gamma$ using high time-resolution, multi-band electromagnetic observations of high-energy astrophysical transients. Typical candidates include blazars \citep{wei2016tests}, the supernova SN1987A \citep{longo1988new}, gamma-ray bursts \citep{gao2015cosmic,nusser2016testing}, fast radio bursts \citep{tingay2016limits,nusser2016testing}, and recently, pulses from the Crab Pulsar \citep{zhang2017test}. A relative time delay $\Delta t_{obs}$ can be interpreted as some combination of factors including intrinsic delay, delays in traversing media along the way, and WEP violation. By assuming there is no fine-tuned cancellation between these, and with a model of the gravitational potential $U(\mathbf{r})$ along the transient's trajectory towards Earth, we can set an upper limit on $\Delta \gamma$ between the two observing wavelengths. This upper limit is saturated when all of the delay is due to WEP violation and still holds if the delay is due to any of the other mechanisms.

Here we present a detailed measurement of the pulse delay between visible and near infrared (NIR) pulses of the Crab pulsar using fast single-photon detectors. Folding $\sim 2\times 10^{5}$ Crab pulses to obtain a high signal to noise ratio enables us to measure the difference in arrival times in these two wavelength bands to within microseconds. We can translate our measurement of the arrival time difference into an upper bound on $\Delta \gamma$ between visible and NIR wavelengths, using the model of the galactic gravitational potential from \cite{zhang2017test}.

\cite{golden2000high,massaro2000fine,rots2004absolute,kuiper2003absolute,oosterbroek2006absolute,molkov2010absolute} have all conducted extensive studies of wavelength dependent Crab pulse delays, measuring the delay with respect to the main radio pulse. Many of these measurements have also been incorporated into the strong upper limits on $\Delta \gamma$ set by~\cite{zhang2017test}. In addition, the Crab pulsar has been timed at optical wavelengths with nanosecond-timing resolution detectors\citep{germana2012aqueye}. However, to the best of our knowledge, our study is the first that combines nanosecond-timing resolution with simultaneous observations in multiple bands to derive an upper limit on violations of the weak equivalence principle as parameterized by $\Delta \gamma$. This work provides a complement to existing bounds from several high-energy astrophysical phenomena, including several from the Crab pulsar, as mentioned earlier.


\section{Observations of the Crab Pulsar}
For our observations, we utilize a custom instrument, described in \cite{leung2017} and similar to those in \cite{handsteiner2017cosmic}, which uses avalanche photodiodes and a time-tagging unit to record the arrival times of single photons from the Crab pulsar with nanosecond-timing resolution in two observing bands simultaneously. We made two observations of the Crab pulsar on the nights of 2016 December 21-22 using the 1-meter telescope at Table Mountain Observatory near Los Angeles, when Crab was at an elevation of $\approx 75^\circ$ in the sky (airmass $X \approx 0.74$ at our altitude above sea level). The first night had clearer observing conditions, but on the second night, we were able to record a 1 pulse-per-second signal from a TM-4 GPS unit, which gave us $\SI{25}{\ns}$ absolute long-term stability referenced to UTC and $10^{-11}$ relative stability over one second. Applying three analyses methods to these two datasets, we obtain new bounds on the variation of the post-Newtonian parameter $\gamma$ between our observing wavelengths, as summarized in Table \ref{tab:observation}.
\begin{table*}
    \centering
    \caption{A summary of our observations, analyses of the pulsar light curves, the resulting upper limits on variations of the post-Newtonian parameter $\Delta \gamma$ between our two observing wavelengths, and a limit on the photon mass. Roughly half of the reported flux is pulsed, with the remaining photons from a combination of skyglow, detector dark counts, and unpulsed flux from the pulsar/nebula. We are also able to derive an upper limit on the photon mass for one of our datasets. See the main text for a more detailed description of our analysis methods.}
    \begin{tabular}{c c c c c c c c}
    \hline 
    \hline
    MJD & Duration $(\si{\second})$ & Flux $(\si{\second^{-1}})$ & Timing & Analysis & $\Delta t_{obs}~(\si{\us})$ & $\Delta \gamma~(3\sigma)$ & $m_{\gamma}~(\si{\gram})$\\
    \hline
    57742 & 6680 & $824$ & Geocentric & Method 1 & $7.41 \pm 0.58$ & $ <1.07\times 10^{-10}$ &  \\
    57742 & 6680 & $824$ & Geocentric & Method 2 & $0.4 \pm 3.6$ &   $ <1.30\times 10^{-10}$ & $<3.7\times10^{-44}$\\
    57743 & 2429 & $763$ & Barycentric & Method 3 & $7.35 \pm 4.48$ & $ <2.4\times 10^{-10} $ & \\
    \hline 
    \end{tabular}
    \label{tab:observation}
\end{table*}

\begin{figure}[ht]
    \centering
    \includegraphics[width = 0.49\textwidth]{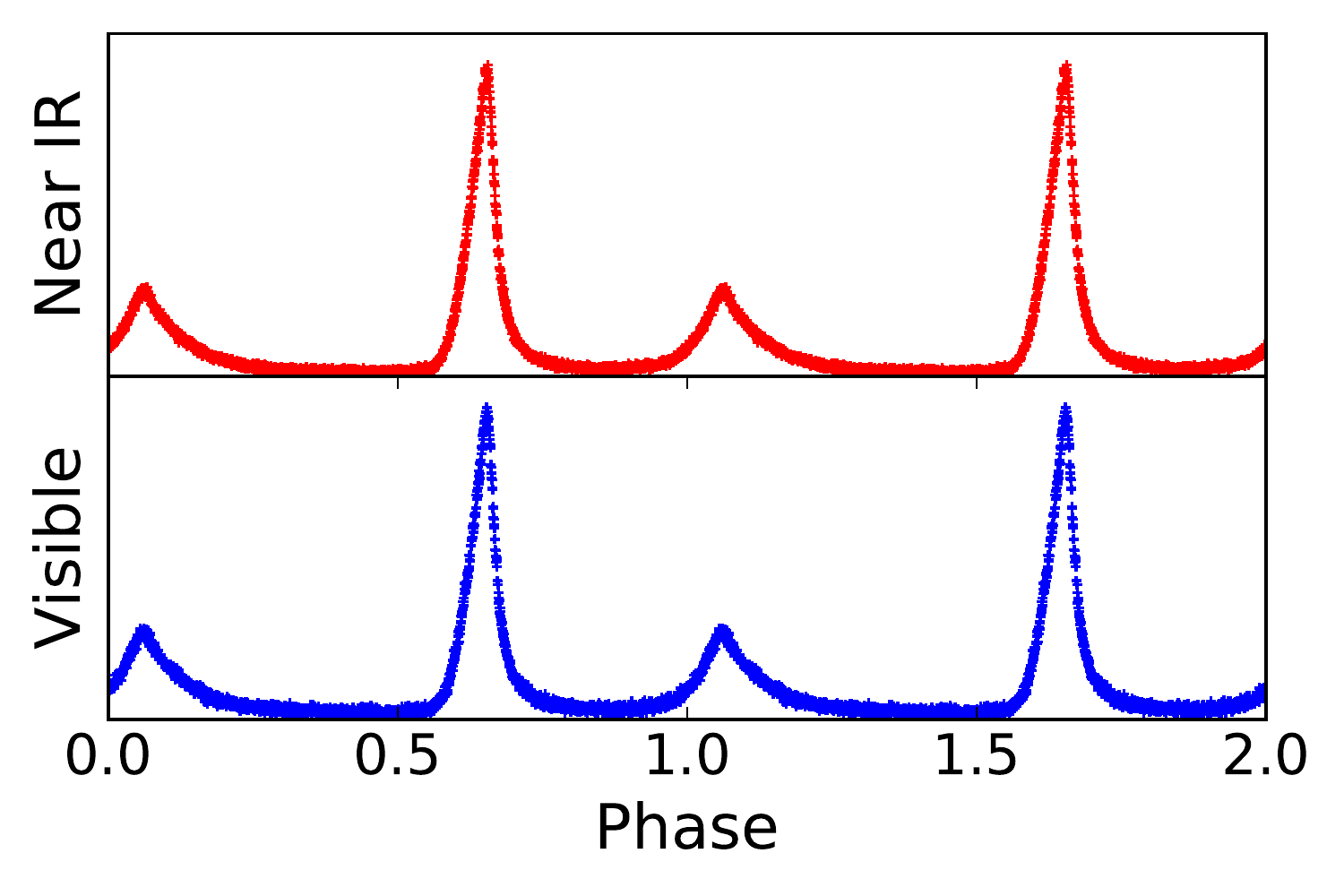}
    \caption{Two periods of our light curves in our near-infrared (top red) and the optical (bottom blue) bands were each constructed from $\sim 3\times 10^{5}$ photon counts. While the signal to noise ratios in each channel are different due to varying skyglow in each channel, we scale the peaks to the same height and subtract the minimum of the light curve to guide the eye. Poisson error bars are too small to show, but they are used in curve fitting and error analysis.}
    \label{fig:rawlightcurves}
\end{figure}

To fit a timing model for the pulsar, we follow a procedure very similar to~\cite{germana2012aqueye}, digitally combining our two lists of detections from the two bands. 
We determine absolute site arrival times by first computing the complex periodogram of our list of $\approx 2\times 10^{6}$ photon detections and maximizing its modulus over possible pulse frequencies $\nu$ to obtain an approximate reference frequency $\nu_{ref} = \SI{29.646635033}{\hertz}$. This reference frequency differs from the pulsar's true rotation frequency by a few parts in $10^5$, primarily due to the Doppler shift induced by the Earth's radial motion towards or away from the pulsar. Folding our data on this reference frequency enables us to compute the light curve of the pulsar, as shown in Fig.~\ref{fig:rawlightcurves}, by binning the photon detections into $1000$ histogram bins. The combined light curves generated are added together and used as a composite template in order to determine site arrival times (SATs) for individual pulsar pulses.

To compute our SATs, we cut our list of photon timestamps into $80$ consecutive observations, each lasting for $\approx\!\!\!\SI{30}{\second}$. We empirically find that this choice balances the trade-off between having more closely-spaced observations and better photon statistics for each observation. Following~\cite{germana2012aqueye}, we convolve our combined visible and NIR template with each individual light curve to determine the value of the phase delay $\Delta \phi$ that maximizes the value of the convolution, thereby defining a (pointlike) SAT for a pulse of nonzero width. Though there exists substantial diversity in the literature with respect to how to do this, with many groups using a least-squares fit of a Lorentzian or a Gaussian function to the pulse's main peak~\citep{kuiper2003absolute,rots2004absolute,oosterbroek2006absolute}, we find in consensus with~\cite{germana2012aqueye} that the aforementioned convolution method is less sensitive than a function fit to the number of histogram bins and is also robust against a fluctuating signal-to-noise ratio. Since the overall shapes of the template and the light curve are very similar, the convolution of the two is to a very good approximation an even function which we notice is very well described by a Lorentzian function. Fitting a Lorentzian to the peak of the aforementioned convolution defines the best-fit delay between the composite template and each individual observation to a statistical precision of $\sim 10^{-4}-10^{-3}$ in pulse phase.

We used {\sc Tempo2}~\citep{hobbs2006tempo2} to convert our SATs into (solar system) barycentric arrival times (BATs), and then fit for the pulsar's frequency and spindown rate. The inputs to {\sc Tempo2} are the GPS coordinates of our telescope, determined to within $\pm 2\si{\m}$ as well as our list of SATs provided in MJD.
We note that cloudy weather obscured the pulsar in several of our 30-second chunks, making peak determination algorithm imprecise due to an poorly-resolved pulse or lack of a pulse altogether. Thus, we post-selected 70/80 of our SATs on the basis of the successful convergence of our peak-finding algorithm to acceptable precision (uncertainty of $<1\times 10^{-3}$ in pulse phase). 

Our $2429$-second observation starting on MJD$=57743.67833$ recorded less than $ 10^{3}$ photons per second from the Crab pulsar. The best-fit pulsar period and period derivative of
\begin{align*} 
1/\nu = P & =0.0337302721 \pm 2.1\times 10^{-10}\si{\second} \\
\dot{P} &= (6.92 \pm 4.2) \times 10^{-13} \si{\second/\second}
\label{eqn:bestpandpdot}
\end{align*}
led to an average pulse arrival time residual of RMS=$\SI{19.9}{\us}$ over our 70 observations. The lack of precision in measuring the spindown $\dot{P}$ is due to our short observation duration.

\section{The Crab Pulsar's Spectrum}
\begin{figure}[ht]
    \centering
    \includegraphics[width = 0.49\textwidth]{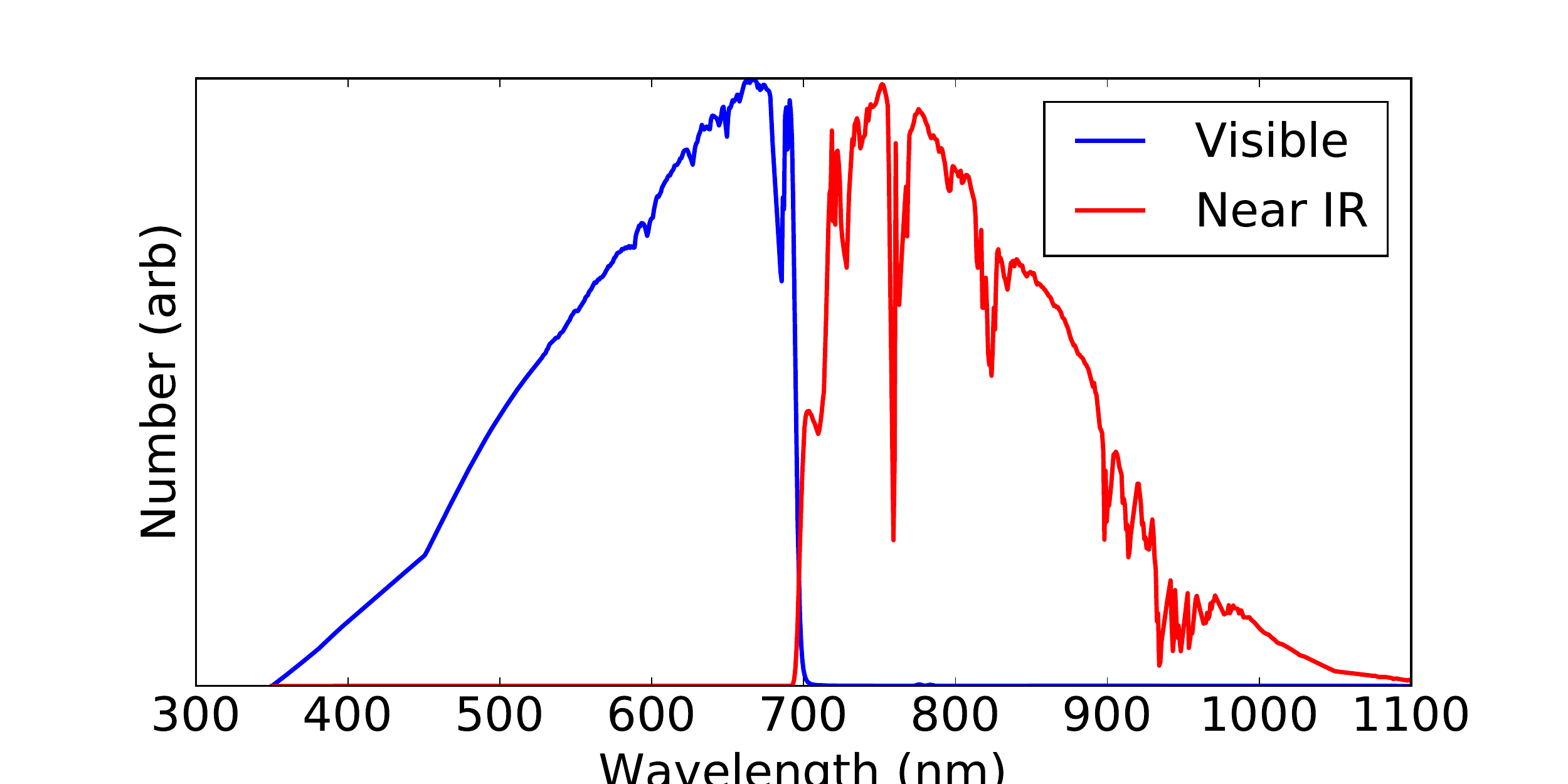}
    \caption{The modeled number distribution of photons detected in our instrument's two observing bands (left blue: transmitted visible arm, right red: reflected NIR arm), from the Crab pulsar after attenuation by atmospheric Rayleigh scattering, optical transmission functions, and our detectors' quantum efficiency curves. Even though our detection channels are closely spaced,  dichroic beamsplitters with steep cutoffs ensure minimal crosstalk between our visible and NIR channels. The probability for a visible photon to be detected in the NIR band and vice versa is $\sim 10^{-3}$. The expected flux ratio between our two bands is in good agreement with our data.}
    \label{fig:crabspectrum}
\end{figure}

We estimate the number distribution of photons detected as a function of wavelength in order to determine our effective observing wavelength. The spectrum of the Crab Pulsar at optical and near-infrared wavelengths is known to follow an empirically-measured power law~\citep{carraminana2000optical}. Using {\sc MODTRAN}\citep{berk1987modtran} to estimate the transmission of the atmosphere due to Rayleigh scattering and telluric absorption, and taking into account the manufacturer-provided transfer functions of each optical component (two achromatic lenses, a detector quantum efficiency curve, as well as a pair of dichroic beamsplitters), we are able to compute up to an overall multiplicative constant the underlying number distribution versus wavelength of the Crab photons detected. Due to the sharp cutoff of our dichroic beamsplitters, the overlap between sensitivity bands is at the level of $\sim 10^{-3}$, minimizing cross-contamination between observing bands. The calculated distribution of detected photons as a function of wavelength arriving in our instrument's two observing bands are plotted in Figure~\ref{fig:crabspectrum}. Photons in our visible and near-infrared bands have average wavelengths and FWHMs of $(585 \pm 180)$ nm, $(815 \pm 177)$ nm respectively. Our data contain $1.01$ near-infrared photons per visible photon, which is roughly consistent with our spectral model which predicts $0.94$.

\section{Delay between Visible and Infrared}
We measure the delay between pulse arrival times in our visible and infrared observing bands with three different methods. 

Method 1 is a point estimate of the delay between the red and blue pulses. This does not incorporate any time dependence or require an absolute timing model of the pulsar. We simply fold our list of detections on our chosen reference period, and determine a ``phase of arrival'' $\phi_{IR}$ by fitting a Lorentzian to the convolution of the infrared light curve with our combined template. Repeating this with our visible light curve gives us $\phi_{vis}$, from which we can determine that the infrared pulse arrives slightly before the visible one, with
$$\Delta t_{obs} = (\phi_{vis} - \phi_{NIR}) / \nu_{ref} = 7.41 \pm 0.58 \si{\us}$$
where the statistical error is estimated by parameter uncertainty on the peak of the Lorentzian fit.

Method 2 is slightly more sophisticated method but also does not require conversion of site arrival times to barycentric arrival times. We break our infrared and visible data into $50$ consecutive observations for each band. We measure the same quantity as in Method 1 in a time dependent way. We determine $\phi_{IR}(t_i)$ and $\phi_{vis}(t_i)$ for each observation taken at time $t_i$ and fit a parabola to the quantity 

$$\phi_{vis}(t) - \phi_{NIR}(t) = at^2 + bt + c.$$

Due to the long-term stability of the pulse profile, we can interpret the constant term $c$ as a phase delay between the blue and red pulses, enabling us to decouple the effects of slow time dependence introduced from uncorrected systematics. We find the best fit parabola 
\begin{align*}
\phi_{vis}(t) &- \phi_{NIR}(t) = (2.1 \pm 3.0)\times 10^{-11} (t-t_0)^2 \\
+&(4.2 \pm 4.8)\times 10^{-8} (t-t_0) + (1.2\pm 10.6)\times 10^{-5}
\end{align*}
where time $t$ is measured with respect to the middle of our observation time $t_0$ in seconds. Note that the quadratic and linear coefficients $a$ and $b$, which absorb time dependent pulse delays, are consistent with zero. The constant term corresponds to a delay of
$$\Delta t_{obs} = 0.4 \pm 3.6 \si{\us}.$$
Method 3 involves using the {\sc Tempo2} timing model described earlier. We input a list of visible SATs and infrared SATs, determined with the same method described earlier, using the same timing model parameters, and use {\sc Tempo2} to fit for a time delay (or a ``jump'') between infrared BATs and visible BATs. We find a delay of 

$$\Delta t_{obs} = 7.35 \pm 4.48 \si{\us}.$$

It is interesting that all of our measurements suggest that the IR pulse arrives before the visible one. However, within our three analysis methods there remains $\sim 2\sigma$ of statistical tension. We suspect that the tension is caused by the pulse's shape differing slightly in our infrared and visible bands, as reported in several works~\citep{percival1993crab,eikenberry1997high,sollerman2000observations}. This systematic effect will be most prevalent in Method 1, which suffers from the least statistical noise. Also, we emphasize that Methods 1 and 2 were carried out with our first dataset with a bad (time tagging module) clock and a good (clear0 sky, whereas Method 3 was only possible using our second dataset with a good clock (GPS 1pps) and a bad (partly cloudy) sky. We hope to resolve this tension with future observations in order to simultaneously obtain precise pulsar timing, as well as more sophisticated data analysis procedures that can possibly mitigate the systematic effect of different pulse shapes.

We characterize our instrumental contribution to this delay to be $(5.2 \pm 0.5) \times 10^{-10}$\,s. 
The largest contribution is relative delay in the two arms of our instrument caused by a small optical path-length difference, and detector/electronics delays. This was quantified by periodically driving a broadband LED at $\sim \SI{700}{\nm}$ to simulate a pulsar's periodic flashing in both of our channels; the total measured latency between the two detection channels was determined to be $(4.8 \pm 0.5) \times 10^{-10}$\,s. In addition, the difference of the index of refraction of air at our two observing wavelengths under our observing conditions is $n(585\si{\nm}) - n(815\si{\nm}) = 2.2 \times 10^{-6}$~\citep{stone2001index}, and corresponds to a relative time delay of $\approx 4.4\times 10^{-11}$ s through our line of sight ($\approx \SI{6}{\km}$) through the atmosphere. Hence the precision of our time delay measurements are not limited by instrumental delays or uncertainty.

\section{Constraining Violations of the Weak Equivalence Principle}

To translate our observed time delays into a measure of WEP violation, we follow recent works ~\citep{longo1988new,gao2015cosmic,wei2015testing,nusser2016testing,tingay2016limits,zhang2017test} in conservatively assuming that the observed delay $\Delta t$ is larger in magnitude than any delay due to a WEP violation. We note that this crucial assumption admits the possibility that a large WEP violation delay is being hidden by an almost equal intrinsic pulse delay in the opposite direction. It is possible to decouple Lorentz-violating WEP delays from astrophysical ones with multiple observations of astrophysical transients coming from different directions~\citep{yu2017robust}, but to our knowledge is not possible to isolate the WEP violation from delays intrinsic to the Crab pulsar or caused by its surrounding nebula for our single observation.

Finally, to calculate $\Delta \gamma$, we employ a model of the gravitational potential $U(\mathbf{r})$ experienced by the pulses as they travel across the Milky Way. We directly employ the model and parameters of \cite{zhang2017test} who formulated it to derive similar limits for the Crab pulsar. The potential $U(\mathbf{r})$ is modeled with two components, a Miyamoto-Nagai disc~\citep{miyamoto1975three} and a Navarro-Frenk-White~\citep{navarro1996structure} dark matter halo, and it is integrated from the Crab pulsar to Earth. This results in a single conversion factor between time delay measurements and $\Delta \gamma$:
$$\Delta \gamma = \SI{1.167e-5}{\s^{-1}} \Delta t_{EPV}$$
For our three analysis methods, we obtain 3$\sigma$ upper limits of 
$$\Delta \gamma_{\mathrm{vis,NIR}} < 1.07\times10^{-10}, 1.30\times10^{-10}, 2.4\times10^{-10}$$
We suspect the large discrepancy between Method 3 versus Methods 1 and 2 is due to the different observing conditions in the two datasets used.

\section{Upper Limit on Photon Mass}
In addition to constraining the WEP, it is possible to obtain a robust upper limit on the photon mass through our observation of a frequency-dependent time delay~\citep{schaefer1999severe}. While global fitting techniques with multiple pulsars can be used to decouple the effects of a nonzero photon mass from plasma dispersion (which induces a similar functional dependence as a function of frequency) ~\citep{wei2018robust}, our measurement of a single pulsar at optical frequencies is largely intrinsically free from that particular systematic effect at $\si{\us}$ levels of precision.

The speed of a massive photon with mass $m_{\gamma}$ is
\begin{equation}
    v/c = \sqrt{1 - \dfrac{m_{\gamma}^2 c^4}{E^2}}
\end{equation}
where the photon energy $E = h\nu$ introduces a wavelength dependence on the speed of light. The observed color-dependent time delay between the arrival times of two photons, to leading order in the observation frequencies $\nu_1 < \nu_2$, is 
$$\Delta t_{obs} = \dfrac{m_{\gamma}^2 c^3 d}{2h^2} (\nu_1^{-2} - \nu_2^{-2}).$$

where $d$ is the light travel distance. For the Crab pulsar, we adopt a distance of $d = 2$kpc for consistency with previous upper limits in~\cite{schaefer1999severe}. It is important to note that this effect, unlike our model-independent test of WEP violation, predicts that bluer photons arrive before redder ones. This corresponds to $\Delta t_{obs} < 0$ which is in statistical tension with two out of our three observations (Method 1 and 3)  which suggest strongly that $\Delta t_{obs} > 0$, refuting altogether the possibility of a photon mass under the crucial assumption that another systematic effect is not canceling the time delay induced by a massive photon. However, we can still derive a quantitative upper limit on the photon mass with our Method 2 result which is consistent within statistical error with a timing measurement of $\Delta t_{obs} \approx 0$; we obtain an upper bound of 

$$m_{\gamma} < \SI{3.7e-44}{\gram}~(3\sigma)$$

under the assumption that $\Delta t_{obs} < 3 \times 3.6\si{\us}$ (the uncertainty of our Method 2 result). While this is not a numerically superlative upper bound on the photon mass, we emphasize that our complementary method of observing at optical frequencies is subject to different systematics and provides a robust check against other limits. For example, recent photon mass constraints using pulsar timing at radio frequencies~\citep{wei2018robust} must carefully remove the systematic effects of plasma dispersion. Plasma dispersion not only exhibits the same functional dependence on observing frequency as a nonvanishing photon mass, but also is estimated by assuming a vanishing photon mass in the first place. In contrast, for our relative timing measurements at optical frequencies to be affected by plasma dispersion at the level of microseconds, the dispersion measure would need to be six orders of magnitude higher than that which is currently measured~\citep{lyne199323}, a highly unlikely possibility. 

\section{Discussion and Conclusion}
\begin{figure}[ht]
    \centering
    \includegraphics[width = 0.49\textwidth]{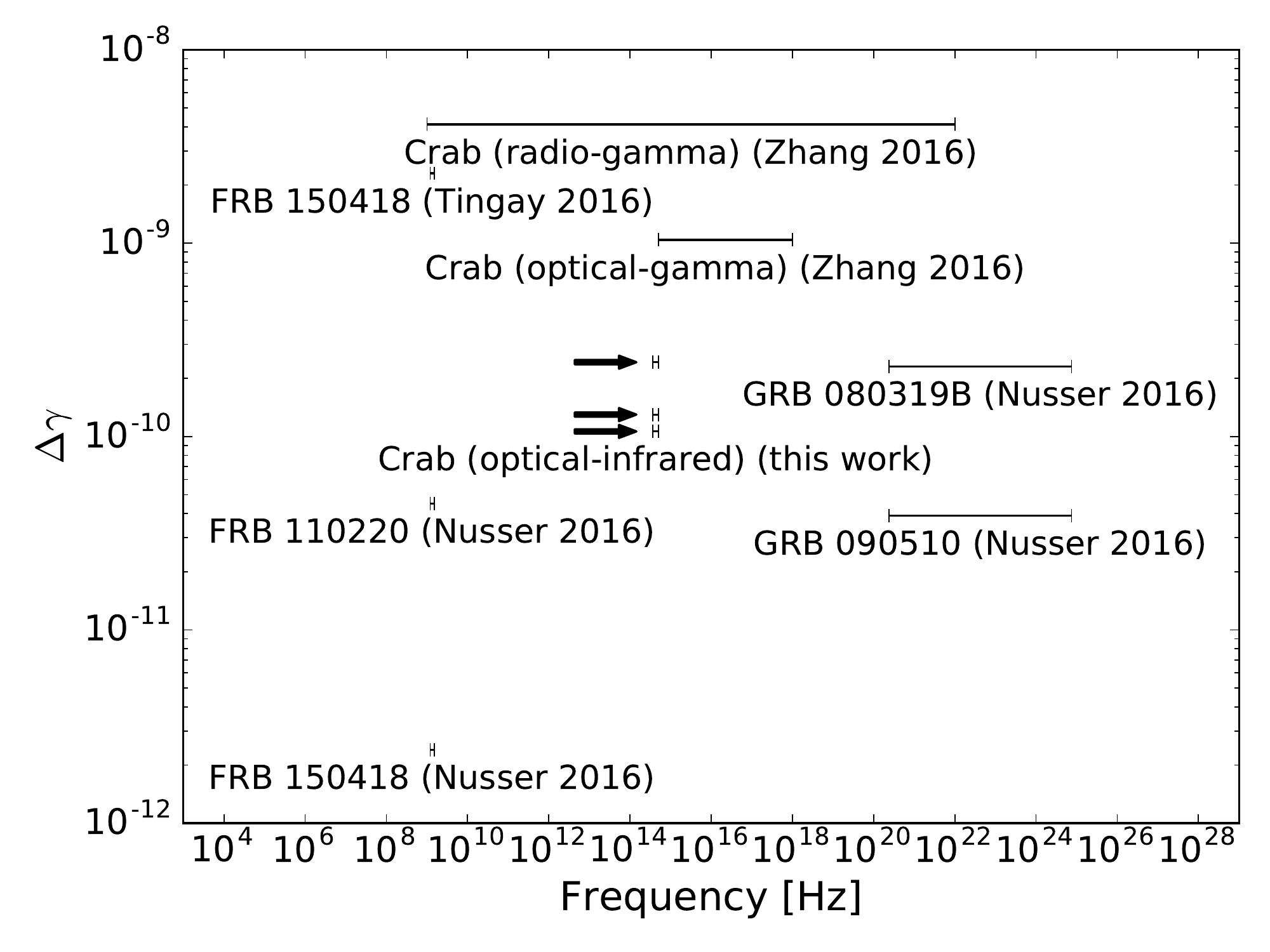}
    \caption{For several astrophysical transients, we plot the most recent and stringent $3\sigma$ upper bounds on values of $\Delta \gamma$ at different observation frequencies across the electromagnetic spectrum. The end points of the error bars correspond to different electromagnetic observings frequencies compared in that particular test. Our results, indicated with arrows, provide a strong complement in an unexplored region of parameter space to existing upper limits from radio and gamma-ray observations.}
    \label{fig:limits}
\end{figure}

Due to our choice of wavelength bands, this upper limit lies in a region of parameter space which makes it complementary to similar photonic tests of WEP violation \citep{wei2015testing,gao2015cosmic,tingay2016limits,zhang2017test}. We summarize $3\sigma$ upper limits from a variety of recent astrophysical tests in Figure~\ref{fig:limits}. 

In this work, we have used a specialized instrument with nanosecond-timing resolution to measure the light curve of the Crab pulsar in two adjacent but disjoint wavelength bands. We first determined the period of the Crab pulsar to sub-nanosecond accuracy with less than an hour-long observation. We set a new upper limit on the differential value of the post-Newtonian parameter $\Delta \gamma$ at visible and near-infrared wavelengths. 

Our upper bound on photonic WEP violation is the first of its kind at the relevant wavelengths. We emphasize that the precision of our measurement is currently limited not by any instrumental uncertainty but rather by the intrinsic ambiguity of assigning a definite value of a delay between two functions with the same period and different profile shapes. The absence of instrumental effects in our equipment at the $\si{\ns}$ level may pave the way to further applications in precise photonic timing measurements. Even though there are only seven known optical pulsars, they can potentially act in tandem with pulsar timing arrays. Unlike radio pulsars, optical pulsar timing measurements are essentially free of plasma dispersion delays. It may be possible to improve our current limits by leveraging the moon's retroreflectors and using strong pulsed lasers at various wavelengths to conduct similar WEP tests.

\section{Acknowledgements}
We are indebted to the support staff at Table Mountain Observatory, particularly Heath Rhodes, for a smooth and productive observing run. Jason Gallicchio acknowledges the support of Harvey Mudd College. Calvin Leung was supported by the Department of Defense (DoD) through the National Defense Science \& Engineering Graduate Fellowship (NDSEG) Program. This research was carried out partly at the Jet Propulsion Laboratory, California Institute of Technology, under a contract with the National Aeronautics and Space Administration and funded through the internal Research and Technology Development program.

%


\end{document}